\documentclass[aps,prb,groupedaddress,twocolumn,longbibliography,superscriptaddress]{revtex4}

\usepackage{graphicx}
\usepackage{amssymb, amsmath}
\usepackage{multirow}
\usepackage{nicefrac}
\usepackage{xcolor}
\usepackage{textcomp}
\usepackage{natbib}
\usepackage[colorlinks=True, allcolors=blue]{hyperref}

\usepackage{pifont}% http://ctan.org/pkg/pifont

\newcommand{\eq}[1]{(\ref{#1})}

\usepackage[normalem]{ulem}

\begin{document}

\title{Theory of coherent optical nonlinearities of intersubband transitions in semiconductor quantum wells}

\author{R.~Cominotti}
\author{H.A.M.~Leymann}
\author{J.~Nespolo}
\affiliation{Pitaevskii BEC Center, CNR-INO and Dipartimento di Fisica, Università di Trento, I-38123 Trento, Italy. }

\author{J.-M. Manceau}
\author{M. Jeannin}
\author{R. Colombelli}
\affiliation{Centre de Nanosciences et de Nanotechnologies (C2N), CNRS UMR 9001, Universit\'e Paris-Saclay, 91120 Palaiseau, France}

\author{I.~Carusotto}
\email{iacopo.carusotto@unitn.it}
\affiliation{Pitaevskii BEC Center, CNR-INO and Dipartimento di Fisica, Università di Trento, I-38123 Trento, Italy. }

\date{\today}

\begin{abstract}
We theoretically study the coherent nonlinear response of electrons confined in semiconductor quantum wells under the effect of an electromagnetic radiation close to resonance with an intersubband transition.
Our approach is based on the time-dependent Schr\"odinger-Poisson equation stemming from a Hartree description of Coulomb-interacting electrons.
This equation is solved by standard numerical tools and the results are interpreted in terms of approximated analytical formulas.
For growing intensity, we observe a red-shift of the effective resonance frequency due to the reduction of the electric dipole moment and the corresponding suppression of the depolarization shift. The competition between coherent nonlinearities and incoherent saturation effects is discussed. The strength of the resulting optical nonlinearity is estimated across different frequency ranges from Mid-IR to THz with an eye to on-going experiments on Bose-Einstein condensation of intersubband polaritons and to the speculative exploration of quantum optical phenomena such as single-photon emission in the Mid-IR and THz windows.
\end{abstract}

\maketitle

\section{Introduction}
\label{sec:intro}
Intersubband (ISB) transitions in semiconductor quantum wells (QW) play a crucial role in a number of opto-electronic devices across a wide range of wavelengths, from the visible down to the IR and the THz ranges~\cite{weber1999intersubband}. A most celebrated example is the quantum cascade laser, which is one of the most widespread semiconductor-based sources of coherent radiation for the Mid-IR and THz ranges of the electromagnetic spectrum~\cite{faist2013quantum}. Nonlinear optical effects in these wavelength regions are also attracting a great interest, in particular for what concerns the realization of passively mode-locked pulsed laser sources~\cite{barbieri-coherent-sampling,modelocking-interband-cascade}, optical combs~\cite{qcl-comb-1, Schwartz-in-out-phase-comb} as well as switching, modulation and harmonic generation when combined with meta-surfaces~\cite{GomezDias_Nonlinear_2015, Lee2014, Mann_Ultrafast_2021}. 

Many among these developments are based on incoherent optical nonlinearities which result from a saturation mechanism due to the shelving of electrons into optically dark states and the consequent reduction of the effective oscillator strength of the transition~\cite{Seilmeier1987,Julien1988,craig1996undressing,Zanotto2015}. 
%{\bf Quote something else} 
%The characteristic time scale of such nonlinearities is set by the decay rate of the dark excitations and is typically in the few-ps range. 
Since they stem from an incoherent dynamics and have a relatively slow time-scale determined by the decay rate of the dark excitations, these nonlinearities can hardly result into coherent processes.
%those coherent light scattering effects like four-wave mixing that are presently under active study in view of ISB polariton lasing or condensation~\cite{Manceau_PRX15}. 
On the other hand, coherent nonlinearities associated to ultrafast processes like Rabi oscillations of a two-level transition has been widely studied under strong pulsed illuminations~\cite{eickemeyer2001coherent,%luo2004nonlinear,
luo2004phase,Dietze_2013}. The goal of this work is to contribute building a general theoretical picture of the nonlinear response of ISB transitions encompassing the two regimes.

The simplest theoretical description of the nonlinear optical response of electrons in quantum wells is based on master equations that only include a few discrete electronic states~\cite{khurgin_coulomb_1991,Dietze_2013, Mann_Ultrafast_2021}. While this is accurate in the low electronic density limit, at higher densities Coulomb interactions start playing a significant role deforming and mixing the single-electron states. At the level of linear response to weak beams, they are responsible for the depolarization shift of the ISB transition~\cite{zaluzny1982inter,ando_electronic_1982,todorov_intersubband_2012,alpeggiani_semiclassical_2014}. For growing light intensities, many intriguing phenomena have been anticipated~\cite{khurgin_coulomb_1991}, in particular the saturation of the optical transition was predicted to give to a corresponding reduction of the depolarization shift and, thus, a sizable frequency shift of the transition~\cite{Zaluzny_NL2}. 
Effects of this kind have been experimentally investigated under quasi-cw illuminations~\cite{craig1996undressing} and theoretically analyzed for the case of  pulsed illuminations~\cite{batista2004rabi}.

While existing approaches are sufficient to obtain accurate predictions for the effective nonlinear optical response in many specific configurations, the objective of this work is to develop a theory that is able to quantitatively capture the coherent dynamics of electrons in generic configurations and include the subtle interplay between nonlinear effects and Coulomb interactions.
Taking inspiration from earlier works~\cite{Zaluzny_NL2,batista2004rabi}, we make use of a mean-field description of the electron dynamics based on a Hartree approximation of Coulomb interactions. This approach yields a time-dependent Schr\"odinger-Poisson equation for the quantum electronic wavefunction in a potential that self-consistently includes the Coulomb interaction as a non-local nonlinear interaction term. 
Here, as a key advance, we numerically solve the time-dependent Schr\"odinger-Poisson equation to obtain predictions for the nonlinear optical response of the electronic system for arbitrary levels of excitation strength and electronic density. 

The structure of the paper is the following. In Sec.~\ref{sec:model} we introduce 
the physical system under investigation and we present our theoretical framework based on the 
Schr\"odinger-Poisson equation. 
This formalism is first applied in Sec.~\ref{sec:linear} to the calculation of the linear optical response 
under weak illumination, recovering, among other, the depolarization shift of the ISB transition 
due to Coulomb interactions at high electronic densities.

In Sec.\ref{sec:ISB} we move on to study the nonlinear optical response to stronger  excitations. In particular we point out a marked intensity-dependence of the ISB resonance and of the effective dipole moment. The numerical results are interpreted in terms of a nonlinear quenching of the oscillator strength and the consequent suppression of the depolarization shift. 
Analytical scaling laws connecting the nonlinear response in different wavelength regions from the Mid-IR to the THz are then extracted from the numerics. A critical discussion of the relation between the coherent nonlinearities and competing incoherent nonlinearities due to the shelving of electrons in dark states is finally provided.

In Sec.~\ref{sec:microcav}, we investigate the consequences of these nonlinearities for a stack of doped QWs embedded in an optical cavity and operating in the strong light-matter coupling regime. Besides the direct red-shift due to the suppressed depolarization shift, in the microcavity case a sizable shift of the polariton branches also occurs from the reduction of the oscillator strength at high intensities. While these two effects are in competition on the lower polariton branch, they add up constructively to reinforce the upper polariton nonlinearity. These predictions are of great importance in the context of intersubband polariton physics with potential application to ISB polariton condensation and lasing~\cite{Manceau_PRX15, nespolo2019generalized}. 
As a final, more speculative topic, in Sec.~\ref{sec:quantum} we discuss the promise of ISB nonlinearities in view of translating to the Mid- and Far-IR domains those polariton blockade effects that were originally predicted for interband exciton-polaritons in the near-infrared and visible range~\cite{verger2006polariton, Delteil_PolaritonBlockade_2019, MunozMatutano_PolaritonCorrelation_2019}, so to explore quantum nonlinear optics phenomena in a novel spectral window.
Conclusions and perspectives are finally drawn in Sec.~\ref{sec:conclu}.

\section{The model and the static properties}
\label{sec:model}

We consider electrons in quantum well systems that are translationally invariant along the $xy$ plane perpendicular to the growth axis $z$. All electrons are assumed to be initially located in the lowest subband of the QW and share the same wavefunction $\psi(z)$ along the growth axis $z$. The total antisymmetry condition of the (fermionic) electronic wavefunction is ensured by the electronic motion along the $xy$ plane, all electronic states of the lowest subband being filled up to a given Fermi energy. Future work will deal with the extension of our formalism to the multi-subband case~\cite{todorov_intersubband_2012,alpeggiani_semiclassical_2014,Deimert__2020} where the electron density is so large that the Fermi energy exceeds the ISB transition energy and the coupled evolution of several wavefunctions corresponding to the different subbands has to be simultaneously determined.

At the level of the Hartree description of Coulomb interactions, we neglect quantum correlations among electrons and we assume that each electron is subject to the electrostatic potential generated by the average electron density. This picture can be formalized in a time-dependent non-local nonlinear Schr\"odinger-Poisson (SP) equation,
%\begin{equation}
 \begin{multline}
i\hbar\frac{\partial\psi(z,t)}{\partial t}=
-\frac{\hbar^2}{2m^*}\frac{\partial^2 \psi}{\partial z^2} +
V_{QW}(z)\psi(z) \\
+\frac{2\pi\, e^2}{\epsilon} \int\!dz'\,|z-z'|\,\left[\sigma_{imp}(z')-\sigma_{el}|\psi(z')|^2\right]\,\psi(z)
\\ - e\,\mathcal{E}(t)\,z\,\psi(z)
\label{eq:NLNLSE}
\end{multline}
%\end{equation}
for the single-particle electron wavefunction $\psi(z)$, normalized to $\int dz\,|\psi(z)|^2=1$. 

$V_{QW}(z)$ is the effective confining potential of the QW and, for simplicity,  we have assumed a parabolic electronic band with a $z$-independent effective mass $m^*$ at all points. Generalization to non-parabolic and/or space-dependent mass case can be carried out with straightforward modifications.
$e$ is the electron charge and $\epsilon$ is the background dielectric constant. Furthermore, $\sigma_{imp}(z)$ is the $z$-dependent three-dimensional density of the dopant impurities and the three-dimensional electronic density is given by $\sigma_{el}\,|\psi(z)|^2$ in terms of the overall surface electron density $\sigma_{el}$. The global neutrality is ensured by $\int dz\,\sigma_{imp}(z)=\sigma_{el}$. The non-locality of the nonlinear interaction term stems from the long-range nature of the Coulomb potential between electrons which, in the present planar geometry, is proportional to the distance $|z-z'|$. In the limit of a vanishing electronic density, the nonlinear interaction term vanishes and the evolution of the single-particle wavefunction $\psi$ reduces to a linear Schr\"odinger equation. Finally, the last term accounts for the time-dependent electric field $\mathcal{E}(t)$ of the electromagnetic wave that excites the electrons. 
%{\bf densite eletronique nulle = terme nonlineaire supprime}

\begin{figure}[ht]
  \centering
  \includegraphics[width=1\columnwidth]{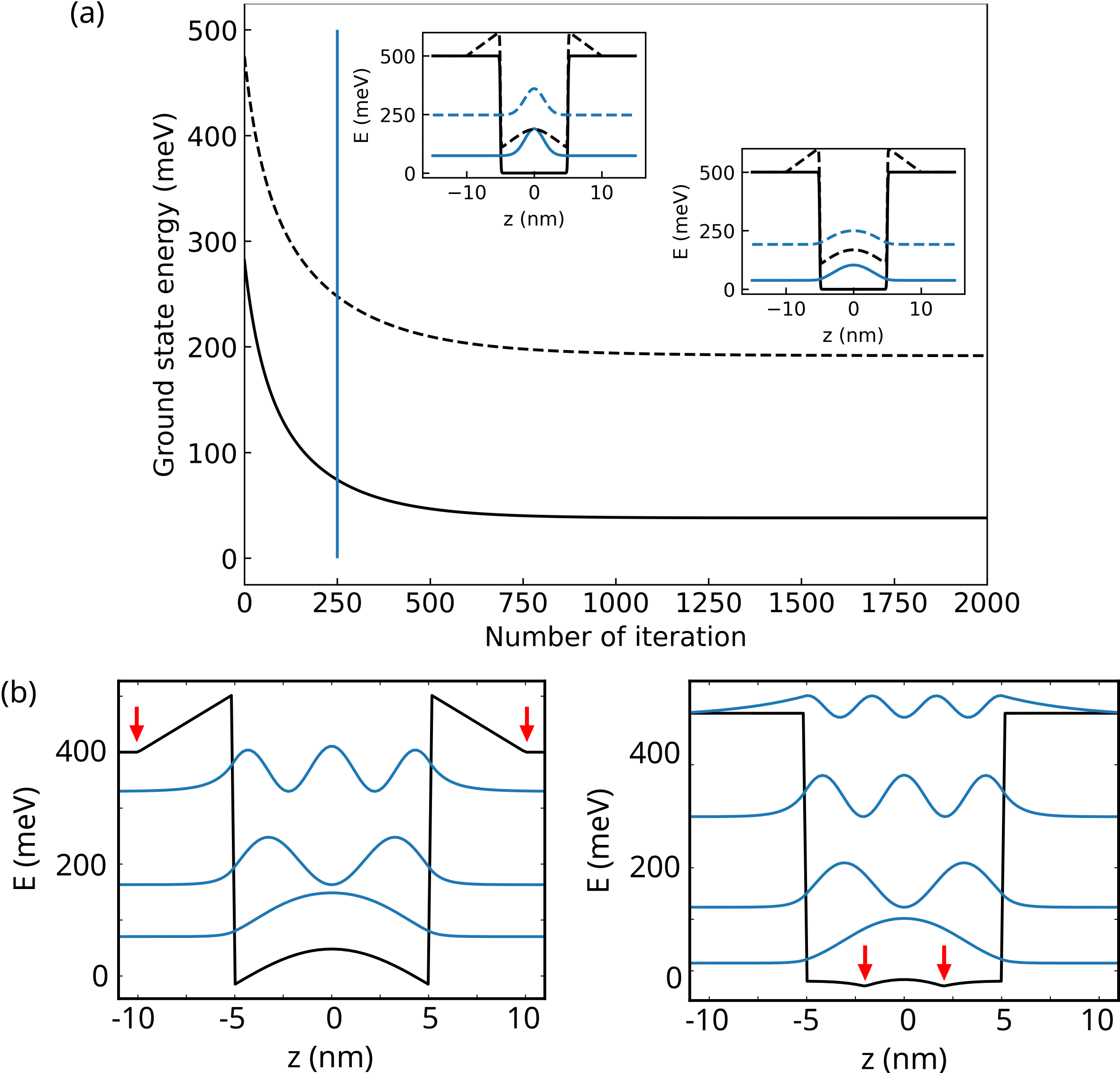}
  \caption{(a) Example of imaginary-time SP evolution of the energy starting from a simply guessed wavefunction (a Gaussian centered on the well) and converging towards the ground state. 
  The insets show the spatial profile of the total potential including the QW and the Coulomb interaction potentials (black lines) and the squared modulus of the electronic wavefunction (blue lines) at different imaginary times. The top inset refers to an early imaginary time indicated by the arrow, well before the convergence of the algorithm. The bottom inset refers instead to a late imaginary time after convergence. Within each panel, the solid/dashed lines refer to different values of the electron density, respectively negligible (solid) and substantial (dashed). 
    (b): Results of the calculation for two different locations of the impurities, as indicated by the red arrows. The black solid line shows the total potential experienced by the electrons. The blue lines show the squared modulus of several single-particle eigenstates for an electron density of $3\cdot 10^{12}$ cm$^{-2}$. Each wavefunction is shifted upwards by an amount corresponding to its eigenenergy. In all calculations within this work, we take values $\epsilon=12.9$ for the background dielectric constant and $m^*=0.067 m_{el}$ for the effective mass of electrons. 
  }
  \label{fig:fig_algo}
\end{figure}

%\section{Static properties and linear dynamics}
%\label{sec:linear}
The first step of our calculation consists of determining the static distribution of charges in the ground state of the system. Because of the Coulomb interaction, the ground state wavefunction is in fact distorted from the lowest eigenstate of the QW potential $V_{QW}(z)$ and can be obtained as the self-consistent minimum energy eigenstate $\psi_g(z)$ of the nonlinear SP equation,
\begin{multline}
E_g\,\psi_g(z)=
-\frac{\hbar^2}{2m^*}\frac{\partial^2 \psi_g}{\partial z^2} +
V_{QW}(z)\psi_g(z)+ \\
+\frac{2\pi\, e^2}{\epsilon} \int\!dz'\,|z-z'|\,\left[\sigma_{imp}(z')-\sigma_{el}|\psi_g(z')|^2\right]\,\psi_g(z)\,.
\label{eq:NLNLSE_stat}
\end{multline}
%of single-particle energy $E_g$.
In order to numerically solve this eigenvalue equation for $\psi_g(z)$ and the single-particle energy $E_g$ in specific configurations, an imaginary-time evolution technique has been adopted. This is a standard numerical technique used to determine the ground state of interacting many-body systems at the level of mean-field approximation, for instance dilute Bose-Einstein condensates of ultracold atoms~\cite{chiofalo2000ground}. A brief account of the principles of this numerical method can be found in Appendix \ref{app:ImT}.

Examples of such calculations are illustrated in Fig.~\ref{fig:fig_algo}. The convergence of the energy to the ground state energy is illustrated in Fig.~\ref{fig:fig_algo}(a) for two different values of the electron density (solid and dashed lines). Snapshots of the electronic wavefunction approaching the ground state are shown as blue lines in the insets of this panel. Here, the black lines display the total potential $V_{\rm tot}(z)$ felt by electrons, equal to the sum of the QW potential $V_{QW}(z)$ and the Coulomb interaction potential. 

Plots of the final wavefunction are shown in Fig.\ref{fig:fig_algo}(b) for the same QW thickness and total electron density $\sigma_{el}=3\cdot 10^{12}\,\textrm{cm}^{-2}$ but different spatial distributions of the dopant impurities, either symmetrically located outside the well (modulation doping, left panel) or inside in the well (right panel), indicated by the red arrows. As before, the solid black lines display the total potential $V_{\rm tot}(z)$ felt by electrons, while the blue lines show the squared modulus $|\psi_i(z)|^2$ of the different eigenstates in the total potential, corresponding to the different subbands. The baseline of each wavefunction is vertically shifted by the energy of each electronic eigenstate. For the sake of simplicity, throughout all this work, we will always assume that the electron population is initially concentrated in the lowest subband.

\section{Linear dynamics}
\label{sec:linear}

\begin{figure}[ht]
  \centering
  \includegraphics[width=1\columnwidth]{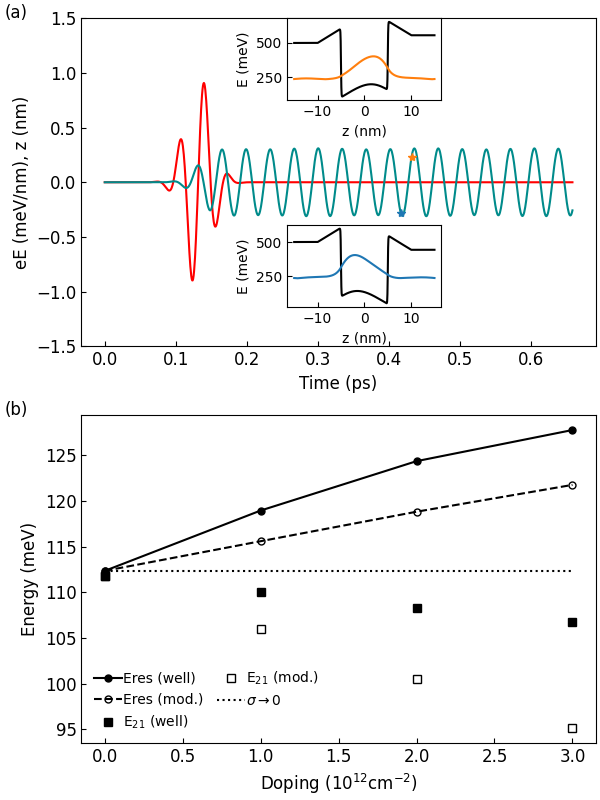}
  \caption{
  (a): Example of the time-dependence of the electric field amplitude of the incident pulse (red) and of the induced electronic polarization (green). The insets show the total potential $V_{\rm tot}(z)$ (black) and the squared modulus of the electronic wavefunction (blue and orange) at two different times indicated by the stars in the main graph. (b) Numerically evaluated ISB transition frequency in the linear regime of weak excitations, as a function of the electron density (solid and dashed lines), compared with the prediction $E_{21}$ of the static Hartree approximation (squares) discussed in the text and in the Appendix \ref{app:Bogo}. The bare transition frequency $E_{21}^o$ is indicated by the horizontal black dotted line.
The dopant impurities are located either outside the well (modulation doping, dashed line and open symbols) or inside the well (solid line and filled symbols. 
  }
  \label{fig:fig_linear}
\end{figure}

The energy differences between the single-particle energy levels shown in the Fig.\ref{fig:fig_algo}(b) give a qualitative indication of the transition energies, but an accurate determination of the collective mode frequencies requires the use of the full SP equation \eq{eq:NLNLSE} since the electronic motion unavoidably leads to a time-dependent modification of the Coulomb potential. In the literature on intersubband transition, this frequency shift goes under the name of depolarization shift~\cite{zaluzny1982inter,ando_electronic_1982,todorov_intersubband_2012,alpeggiani_semiclassical_2014}. 

An exact prediction for the collective oscillation frequencies in the linear regime of weak excitations could be obtained by linearizing the SP equation \eq{eq:NLNLSE} around the ground state $\psi_g$ determined in the previous Section. This gives the linearized equation of motion
\begin{multline}
i\hbar \frac{\partial\,\delta\psi(z,t)}{\partial t}=-\frac{\hbar^2}{2m^*}\frac{\partial^2 \delta\psi(z)}{\partial z^2} +
V_{QW}(z)\,\delta\psi(z) \\
+\frac{2\pi\, e^2}{\epsilon} \!\int\!dz'\,|z-z'|\,\left\{\left[\sigma_{imp}(z')-\sigma_{el}|\psi_g(z')|^2\right]\,\delta\psi(z)+ \right.\\\left.-\sigma_{el}\left[\delta\psi^*(z')\,\psi_g(z')+\psi^*_g(z')\,\delta\psi(z')\right]\psi_g(z) \right\}
\label{eq:Bogo_t}
\end{multline}
for the perturbation $\delta\psi(z)$ defined as $\psi(z)=e^{-iE_g t/\hbar}[\psi_g(z)+\delta\psi(z,t)]$. Differently from the Bogoliubov equations for atomic gases with contact interactions~\cite{pitaevskii2016bose,castin2001bose}, here the linearized equation features a nonlocal Coulomb interaction term. 

An analytical solution for the eigenmodes of \eqref{eq:Bogo_t} can be obtained only in suitable limits as discussed in Appendix \ref{app:Bogo}. In the general case, it typically requires numerical methods to diagonalize the linear problem~\footnote{The typical strategy to deal with the $\delta\psi^*$ term in \eqref{eq:Bogo_t} is to consider $\delta\psi$ and $\delta\psi^*$ as independent variables and solve the system of two equations describing their coupled dynamics~\cite{pitaevskii2016bose,castin2001bose}.}. For this reason, in this work we adopt an equivalent, yet fully numeric strategy that has the key advantage of being directly extended to the nonlinear regime.
As it is sketched in Fig.\ref{fig:fig_linear}(a), the idea is to simulate the real-time evolution starting from the ground state wavefunction $\psi_g(z)$ obtained by the imaginary-time evolution discussed in the previous Section and look at the small oscillations induced by a suitably chosen weak electromagnetic pulse $\mathcal{E}(t)$. 

The carrier frequency of the pulse is chosen to be in the vicinity of the transition frequency of interest. The pulse duration is chosen to be long enough in time not to excite multiple excitation modes, but also short enough to give a sufficiently wide bandwidth that easily covers the desired transition. As a rule of thumb, Gaussian pulses with a duration in the several 10~fs range are typically a good choice to efficiently excite ISB transitions in the Mid-IR without having to fine-tune the carrier frequency. 
The overall pulse strength is chosen to be weak enough to be well in the linear regime and avoid (for the moment) all nonlinear effects. 

We then record the temporal evolution of the system at later times in response to the perturbation. As a key observable, we calculate the time-dependence of the average electronic polarization
\begin{equation}
    z(t) = \int \! dz\,z\,|\psi(z,t)|^2\,.
\end{equation}
An example of the temporal profile of the applied electric field pulse is shown as a red line in Fig.\ref{fig:fig_linear} together with the resulting time-dependence of $z(t)$ (blue curve). From this latter, the transition frequency $\omega_{\rm res}$ and the oscillation amplitude $\bar{z}$ are straightforwardly obtained by means of a sinusoidal fit of the form~\footnote{Alternatively, the oscillation frequency $\omega_{\rm res}$ can be extracted from the position of the dominant peak in the Fourier transform of $z(t)$ at late times. Both these strategies give equivalent results and, most importantly, are completely independent of the details of the specific initial excitation pulse used in the calculation. } 
\begin{equation}
    z(t)\simeq \bar{z}\,\cos(\omega_{\rm res} t +\varphi)\,.
\end{equation} 
%or, alternatively, by looking for the dominant peak in the Fourier transform of $z(t)$. 
 Note that the fact that in our calculation the oscillations persist for indefinitely long times is a consequence of having neglected all the electronic decoherence processes~\cite{Unuma2003} that would lead to a decay of the dipole moment at a characteristic rate $\gamma_{ISB}$ typically on the order of a fraction of ps. Inclusion of decoherence processes into our Schr\"odinger-Poisson formalism will be the subject of future work. 

The result of a few such calculations for realistic QW parameters are shown in Fig.\ref{fig:fig_linear}(b), where the oscillation frequency $\omega_{res}$ in response to a very weak perturbation in the linear regime is plotted as a function of the electron density for dopant impurities {located outside the well (modulation doping) or inside the well}.
For small electron densities, the Coulomb interactions have a negligible effect and the oscillation frequency recovers the energy separation %$\Delta_{21}^o$ 
$E_{21}^o$ of the two lowest eigenstates of the linear Schr\"odinger equation in the bare potential $V_{QW}(z)$ of the well. 

The significant blue-shift of the oscillation frequency that is observed for growing electron densities %(rhombus, pentagons and circles) 
is a result of Coulomb interactions. Here, the importance of the depolarization shift~\cite{zaluzny1982inter,ando_electronic_1982,todorov_intersubband_2012,alpeggiani_semiclassical_2014} is easily appreciated comparing the numerically obtained oscillation frequency (solid and dashed lines) with the static Hartree prediction given by the energy difference $E_{21}$ %$\Delta_{12}$
between the two lowest energy single-particle orbitals in the total static potential $V_{\rm tot}$ including interactions (square symbols)~\footnote{These are obtained as the eigenfunctions of the effective Schr\"odinger equation in the static potential. Of course, the lowest energy such eigenstate corresponds to the ground state obtained by imaginary-time evolution.  }. Additional analytical insight on the physical origin of this frequency difference is presented in Appendix \ref{app:Bogo}.

\section{Nonlinear dynamics}
\label{sec:ISB}

In the previous Section we have introduced our theoretical framework and we have given a first confirmation of its efficiency by numerically recovering the well-known depolarization shift of the oscillation frequency in the weak excitation regime. In the present Section we move on to the study of nonlinear effects. Among the many nonlinear effects that ISB may feature in different configurations, in this work we focus on the shifts of the oscillation frequency that appear under stronger excitations and on the related drop of the oscillator strength. The numerical results will be interpreted in terms of simple analytical relations and scaling laws that allow to quickly bridge different regimes and wavelength ranges. The experimental consequences of our results will be finally discussed in comparison with competing incoherent effects.

\subsection{Numerical results}

Nonlinear effects start being visible for larger values of the applied electric field amplitude. As done in the previous Section for the linear regime, the oscillation frequency $\omega_{\rm res}$ and the oscillation amplitude $\bar{z}$ are extracted from the sinusoidal fit of the time-dependent dipole $z(t)$ at late times. % or, equivalently, from its Fourier transform. %The excitation energy is estimated from the total (kinetic plus potential plus interaction energy) of the electrons after the pulse has ended.
Once again, provided the excitation pulse spectrum is narrow enough to only excite the desired ISB transition, we have verified that the extracted value of $\omega_{res}$ and $\bar{z}$ do not depend on the details of the excitation pulse but only on the energy that gets deposited in the QW as expected on physical grounds.

As a crucial step to unravel the different mechanisms underlying the nonlinear effects, we need to isolate the intrinsic nonlinear dependence of the oscillation frequency on the excitation level from the nontrivial dynamics of the excitation process during the electric field pulse. As it was pointed out in Ref.~\onlinecite{batista2004rabi}, the amount of energy that gets effectively delivered to the electrons may in fact be strongly affected by nonlinear effects that push the transition frequency on- or off-resonance during the pulse itself.

\begin{figure}[htbp]
  \centering
  \includegraphics[width=1\columnwidth]{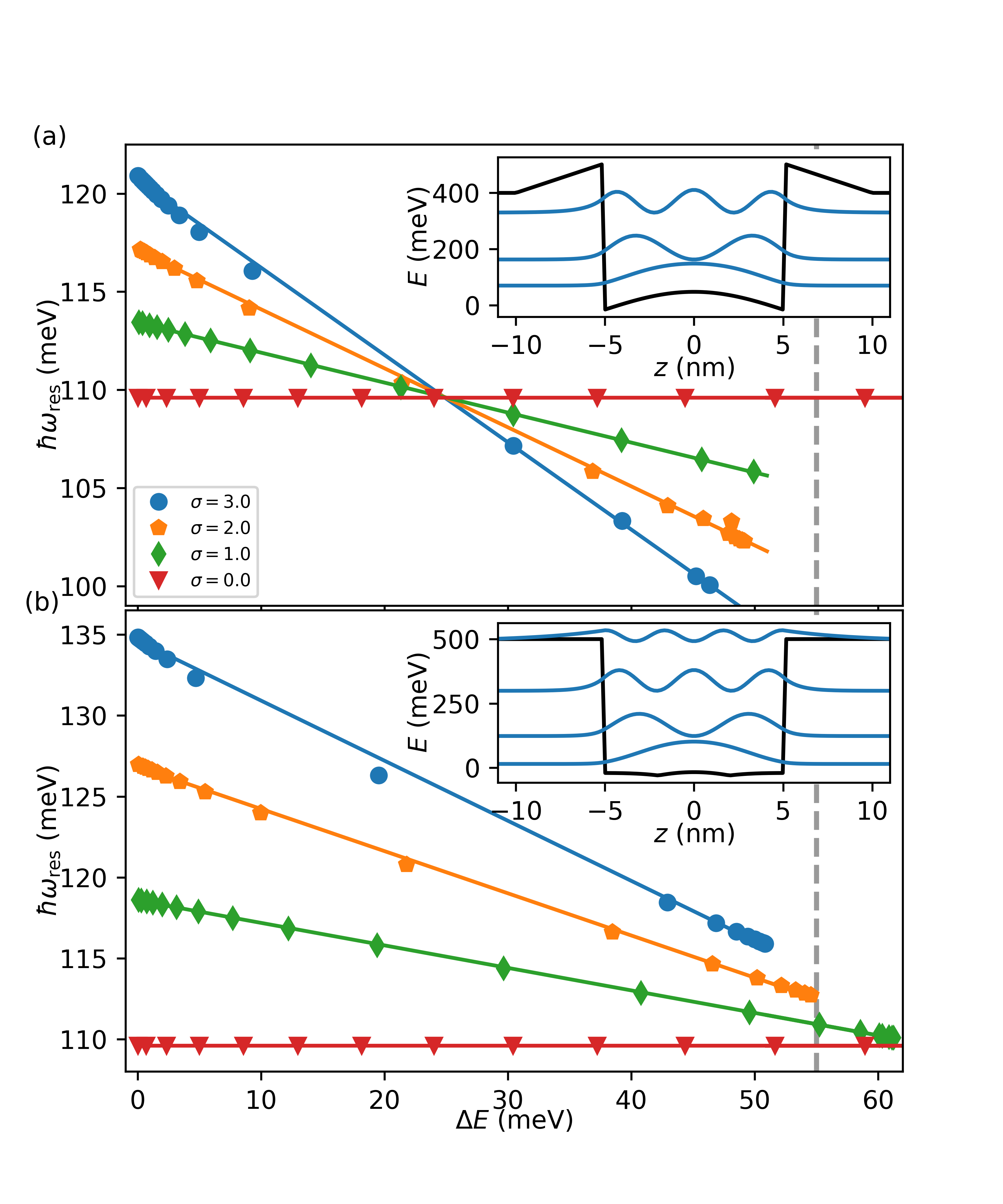}
  \caption{
Oscillation frequency $\hbar\omega_{\rm res}$ as a function of the deposited energy per electron $\Delta E$. Different curves refer to growing values of the electron density, from a very small $\sigma_{el} \sim 0\,\textrm{cm}^{-2}$ value (red triangles), to higher ones $\sigma_{el} = 1\cdot 10^{12}\,\textrm{cm}^{-2}$ (green diamonds), $2\cdot 10^{12}\,\textrm{cm}^{-2}$ (yellow pentagons) and $3\cdot 10^{12}\,\textrm{cm}^{-2}$ (blue dots). The two (a,b) panels refer to different location of the dopant impurities outside (a) and inside (b) the QW. Solid lines are guides to the eyes based on a linear regression of the numerical data points. Within each panel, the inset shows (black line) the total potential $V_{\rm tot}(z)$ in the ground state and (blue lines) the squared modulus of several single-particle eigenstates in the total potential for the highest value of the electronic density considered in the main panels. Each wavefunction is shifted upwards by an amount corresponding to its eigenenergy. %In all calculations of this work, we take values $\epsilon=12.9$ for the background dielectric constant and $m^*=0.067 m_{el}$ for the effective mass of electrons. 
 For the chosen QW parameters, the bare ISB transition is in the Mid-IR range at an energy around $110$~meV. The vertical dashed lines indicate the position of the saturation point $\Delta E=E_{21}^o/2$. 
}
  \label{fig:energyres}
\end{figure}

To suppress these effects, we classify our numerically extracted value of the oscillation frequency as a function of the additional energy  (per electron) $\Delta E$ that is deposited in the electronic motion at the end of the excitation sequence. Once the pulse has gone, energy is conserved, so $\Delta E$ can be estimated by numerically calculating the increase in the SP energy
\begin{multline}
E_{SP}=\int\!dz\,\left[
\frac{\hbar^2}{2m^*}\left|\frac{\partial \psi}{\partial z}\right|^2 +
V_{QW}(z)\,|\psi(z)|^2 \right]+ \\
+\frac{\pi\, e^2}{\epsilon} \int\!dz\,dz'\,|z-z'|\,\left[\sigma_{imp}(z')-\sigma_{el}|\psi(z')|^2\right]\,|\psi(z)|^2
\label{eq:energy_SP}
\end{multline}
between the wavefunction $\psi$ at a generic late-time $t_{\rm late}$ and the one at the initial time $t=0$, $\Delta E=E_{SP}(t_{\rm late})-E_{SP}(t=0)$. 
An intuitive understanding of the physical meaning of the $\Delta E$ parameter can be obtained by relating it to experimentally accessible quantities. For sufficiently weak Coulomb interactions, we can neglect nonlinearities and approximately write $\Delta E= p_2\,E_{21}^o$, where $p_{1,2}$ are the fractional occupations of the ground and excited subband, $p_1+p_2=1$. The point $\Delta E=E_{21}^o/2$ then corresponds to the saturation condition $p_2=p_1=1/2$. Whereas these arguments are exact in the low-density limit, they remain qualitatively valid also in the presence of sizable nonlinearities as long as Coulomb energy is a relatively small correction to the bare transition energy $E_{21}^o$. Additional remarks on the relation between $\Delta E$ and the incident intensity in a microcavity geometry will be given at the beginning of Sec.\ref{sec:microcav}

\subsubsection{Nonlinear shift of the oscillation frequency}
Fig. \ref{fig:energyres} shows the oscillation frequency $\omega_{res}$ as a function of the deposited energy $\Delta E$ for two different spatial distributions of the dopant impurities, namely outside (top) or inside (bottom) the well: as expected from the approximated calculations in Ref.~\onlinecite{Zaluzny_NL2}, the oscillation frequency $\hbar\omega_{res}$  shows an approximately linear red-shift for growing $\Delta E$. 

For a given geometry, we notice from this figure that the different lines corresponding to different values of the electron density $\sigma_{el}$ approximately cross at a single point on the $\hbar \omega_{\rm res}=E_{21}^o$%=\Delta_{21}^o$ 
horizontal line. Together with the linear dependence of the linear-regime depolarization shift on $\sigma_{el}$ discussed above, this points towards an approximated analytical form
\begin{equation}
 \frac{\hbar \omega_{\rm res}}{E_{21}^o}\simeq 1 + \frac{a_0\, e^2 L_{QW} \sigma_{el}}{\epsilon\,E_{21}^o} -\frac{a_1\, e^2 L_{QW} \sigma_{el}}{\epsilon\,E_{21}^o} \frac{\Delta E}{E_{21}^o}
 \label{eq:shift_omega}
\end{equation}
%{\bf rajouter expression en termes de $\omega_{plasma}=4\pi e^2 \sigma/m^*$ et comparer avec depolarization $\sqrt{\omega_{12}^2+\omega_{plasma}^2}$}
for the resonance frequency, where the adimensional $a_{0,1}$ parameters only depend on the well geometry and not on the electron density. It is interesting to note how the second term in \eqref{eq:shift_omega} has a similar functional form as the standard expression for the depolarization shift 
\begin{equation}
    \frac{\hbar\omega_{res}}{E_{21}^o}=\frac{\sqrt{{E_{21}^o}^2 + (\hbar\omega_{pl})^2}}{E_{21}^o} \simeq 1 + \frac{\hbar^2\omega_{pl}^2}{2 {E_{21}^o}^2}
\end{equation}
where $\omega_{pl}=[4\pi e^2 \sigma_{el} / (m^* \epsilon L_{QW})]^{1/2}$ is the plasma frequency of QW electrons and the $a_0$ coefficient in \eqref{eq:shift_omega} is related to the adimensional geometrical parameter $\hbar^2/(m^* L_{QW}^2 E_{21}^o)$ of order one.

For the configurations of Fig.\ref{fig:energyres}, the $a_{0,1}$ parameters are approximated by $a_0\sim 0.35$ and $a_1\sim 1.5$ for the top panel and by $a_0\sim 0.6$ and $a_1\sim 1.3$ for the bottom one. As it is discussed in detail in Appendix \ref{app:Bogo}, the smaller value of the weak-excitation frequency shift parameter $a_0$ in the upper panel is mostly the effect of the static Coulomb interactions with the doping impurities, which are more significant when these latter are located outside the well and counterbalance the effect of the dynamical distortion of the electronic distribution induced by the excitation.   
On the other hand, the slope $a_1$ of the excitation-dependent shift has comparable values in the two cases as it mostly depends on depolarization shift effects. Quite interestingly, for the experimentally realistic values of the electron density considered in the figure, the nonlinear red shift can be comparable to the typical linewidths of ISB transitions (around a few meV's) already for deposited energies well below saturation of the two-level ISB transition, that is $\Delta E \ll E_{21}^o/2 \simeq 55$\,meV (vertical dashed line in the Figure).  As we are going to see in the next Subsec.\ref{sec:incoh}, this statement may need revisiting if a sizable fraction of the electrons gets accumulated in the QW as incoherent dark excitations.

\begin{figure}[ht]
  \centering
  \includegraphics[width=1\columnwidth]{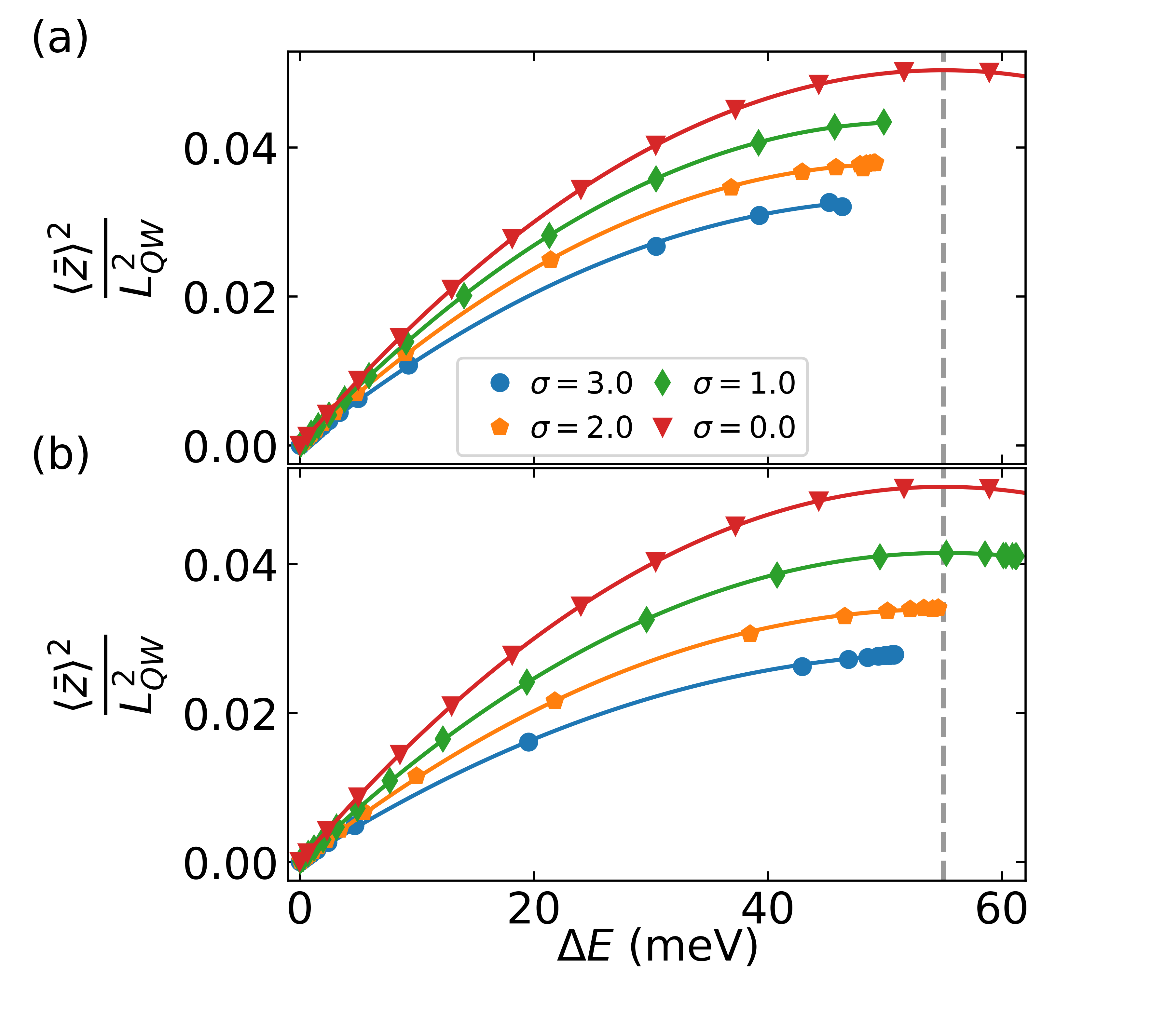}
  \caption{Normalized squared amplitude of the oscillating electric dipole $\bar{z}$ as a function of the deposited energy $\Delta E$ for the same configurations shown in Fig.\ref{fig:energyres}. The vertical dashed lines indicate the position of the saturation point $\Delta E=E_{21}^o/2$. 
  }
  \label{fig:energydip}
\end{figure}

\subsubsection{Oscillation amplitude}

Similar plots for the amplitude $\bar{z}$ of the electron oscillations as a function of the excitation energy are shown in Fig.\ref{fig:energydip} for the same well geometries and electron densities as in the previous figure. Physically, $\bar{z}$ corresponds to the effective electric dipole moment of the transition, a quantity closely related to its oscillator strength.

The general features of this plot are easily understood in the case of a negligible electron density (red triangles). In this case, one recovers the usual behaviour of two-level atoms. The squared dipole moment $\bar{z}^2$ grows linearly for small excitation energies $\Delta E$, then saturates to a maximum value for $\Delta E=E_{21}^o/2=55$~meV where the population is equally distributed between the two levels, and finally decreases again in the population-inversion regime. 

Except for an overall reduction of the dipole moment due to the modification of the electronic wavefunctions by Coulomb interactions, a qualitatively very similar behaviour is found for higher electron densities $\sigma_{el}$, with a linear growth of $\bar{z}^2$ at small $\Delta E$ and a saturation at larger $\Delta E$. For all geometries and all values of the electron density considered, our simulations confirm the physical expectation that the transition gets saturated when approximately half of the electronic population is in the excited states of the well. 

This behaviour can be summarized in an approximated analytical form
\begin{equation}
\left( \frac{\bar{z}}{L_{QW}}\right)^2 \simeq b_0^2\,\left[1- 
 b_1\frac{\Delta E}{E_{21}^o}\right]\,\left[1-s_1\frac{e^2 \sigma_{el} L_{QW}}{\epsilon E_{21}^o}\right]\,\frac{\Delta E}{E_{21}^o},
\label{eq:z0}
\end{equation}
where the adimensional $b_{0,1}$ and $s_1$ coefficients only depend on the well geometry. % and have the values... for the configurations considered in Fig.\ref{fig:energydip}. 
In all considered cases, from the data displayed in the Fig.\ref{fig:energydip} one can extract a value $b_1\sim 1$ for the coefficient in the nonlinear dipole moment reduction factor, which gives a factor $1/2$ suppression of the squared dipole moment when $\Delta E/E_{21}^o\simeq 1/2$ and electrons are equally (yet coherently) distributed between the two subbands.
This is in stark contrast with the case of incoherent nonlinearities, where an equal population of the two states leads to a  full quenching of the dipole moment~\cite{Seilmeier1987,Julien1988,craig1996undressing,Zanotto2015}. % {\bf add some reference here}. 
%Note that this value is in rough agreement with the $a_1/a_0 \sim 0.58$ ratio extracted from the bottom panels of Figure 1 and quantifying the excitation-induced reduction of the depolarization shift. For the upper panel, the difference to $a_1/a_0\sim 0.22$ is much more pronounced since the shift of the transition energy by the static Coulomb interactions is more important.

%As expected from~\cite{Zaluzny_NL2}, the coefficient $b_1$ describing the saturation of the oscillator strength is close to half the coefficient $a_1/a_0$ describing the suppression of the depolarization shift in \eqref{eq:shift_omega}. CHECK!

\subsection{Scaling laws}
The coefficients in the analytical forms \eqref{eq:shift_omega} and \eqref{eq:z0} have an interesting interpretation in terms of scaling laws. For the purpose of this discussion, let us restrict for simplicity to quantum wells of thickness $L_{QW}$ with infinitely high barriers. As compared to the more realistic configuration considered in the Figures, having infinite barriers only introduces quantitatively minor differences.

If lengths are measured in units of the well thickness $\zeta=z/L_{QW}$ and, correspondingly, time in units of the inverse kinetic energy $\tau=\hbar t /(m L_{QW}^2)$,
the Schr\"odinger-Poisson equation can be recast in an adimensional form
\begin{multline}
i\frac{\partial \tilde{\psi}}{\partial \tau} = -\frac{1}{2}\,\frac{\partial^2 \tilde{\psi}}{\partial \zeta^2} + \tilde{V}(\zeta)\,\tilde{\psi}-\tilde{\mathcal{E}}(\tau) \, \zeta\, \tilde{\psi} + \\ 
%+\frac{2 \pi e^2 m L_{QW}^3}{\hbar^2 \epsilon} 
+\eta_{Coul}
\int \! d\zeta\,|\zeta-\zeta'|\,[\tilde{\sigma}_{imp}(\zeta')-\sigma_{el}\,|\tilde{\psi}(\zeta')|^2 ]\,,
\end{multline}
where the renormalized wavefunction $\tilde{\psi}=L_{QW}^{1/2} \psi$ is normalized such that $\int d\zeta\,|\tilde{\psi}(\zeta)|^2=1$, the renormalized impurity density is $\tilde{\sigma}_{imp}= \sigma_{imp} L_{QW}$, and electric field is  $\tilde{\mathcal{E}}= e \mathcal{E} m L_{QW}^3/\hbar^2$. The function $\tilde{V}(\zeta)$ describes the infinite barriers and is defined as $\mathcal{V}=0$ for $|\zeta|<1/2$ and $+\infty$ otherwise. The relative strength of the Coulomb interactions is quantified by the coefficient
$\eta_{Coul}={2 \pi e^2 m L_{QW}^3} \sigma_{el}/{\hbar^2 \epsilon}$ on the second line.

The universality of this form provides simple scaling laws under which the coefficients in Eqs.\eqref{eq:shift_omega} and \eqref{eq:z0} are invariant. This provides a straightforward way to extend our analytical results to all frequency ranges without the need of repeating the numerical calculation. Specifically, let us consider that the QW width is varied by a factor $\alpha^{-1/2}$ so that the resonance frequency $E_{21}^o$ is multiplied by $\alpha$. Under this change, the Coulomb interaction parameter $\eta_{Coul}$ (i.e. the relative value of the depolarization shift) stays constant if the electron density is varied by a factor $\alpha^{3/2}$. 
For instance, reducing  the  resonance  frequency  by  a  factor  10  from  110~meV (in the Mid-IR) to  11~meV  (corresponding to $\sim 2.7$~THz) requires a $\sim 3$ times wider well; keeping the same $\eta_{Coul}$ then requires a $\sim 30$ times lower two-dimensional electron density. In the next Section, we will see how this scaling impacts the value of light intensity that is needed to observe nonlinear effects.

\subsection{Competing incoherent nonlinearities}
\label{sec:incoh}
%\section{Incoherent saturation} 

Our theoretical developments so far provide a simple, yet realistic model of those coherent optical nonlinearities that stem from the intrinsic nonlinearity of the electronic motion and the Coulomb interactions. As such, our results directly provide an estimate of the magnitude of coherent nonlinear processes, for instance the nonlinear frequency shift of the ISB resonance or, equivalently, the strength of the parametric coupling in wave-mixing processes. %is directly related to the nonlinear frequency shift studied in the previous subsections. 

It is however crucial to keep in mind that other nonlinear processes of incoherent nature are also typically at play for intersubband transitions in QWs.
Electrons in QWs are in fact subject to different decoherence mechanisms that lead to a fast effective decay of the intersubband excitations into relatively long-lived dark electronic excitations~\cite{Unuma2003}. This results in a sizable reduction in the density of active electrons participating to the electronic transition and thus in a quenching of the oscillator strength and, as pointed out in early works~\cite{Zaluzny_NL2}, of the depolarization shift. 

Leaving aside the shift of the transition frequency due to the static Coulomb interactions with the dark excitations (which is typically  small in simple wells, but may become sizable in strongly asymmetric configurations~\cite{khurgin_coulomb_1991}), the coherent optical response can still be captured by our theory provided we identify at each time the electron density with the one of active electrons, $\sigma_{el} \rightarrow \sigma_{el}^{\rm act}$ and we estimate the evolution of $\sigma_{el}^{\rm act}$ in time using a simple rate-equation model,
\begin{equation}
    \dot \sigma_{el}^{\rm act}= \gamma_d (\sigma_{el}-\sigma_{el}^{\rm act}) - \gamma_{ISB} \sigma^{\rm act}_{el} \frac{\Delta E}{E_{21}^o}
\end{equation}
where $\gamma_{ISB}$ is the decay rate of the coherent ISB excitations (of density approximately given by $\sigma_{el}^{\rm act}\,\Delta E/E_{21}^o\simeq p_2 \sigma_{el}^{\rm act}$) and $\gamma_d$ is the (typically much slower) decay rate of the dark excitations. Estimates for this latter are typically in the 10~ps range, much longer than the characteristic decay of the coherent ISB excitation on the order of a fraction of ps. % $\gamma_d^{-1}\sim 5$ to $100~$ps 

%{\bf discuter s'il en reste assez sur CW pour donner oscillation parametrique. Discuter aussi aue $I_{sat}$ croit avec $\sigma$, tandis que la nonlinearite coherente ne croit pas. Rappel de Belkin.}
At steady-state under a monochromatic excitation, the fraction of active electrons is reduced to
\begin{equation}
\frac{\sigma_{el}^{\rm act}}{\sigma_{el}}=\frac{1}{1+\frac{\Delta E}{E_{21}^o}\frac{\gamma_{ISB}} {\gamma_d}}\,.
\label{eq:el_act}
\end{equation}
Since in typical samples $\gamma_d\ll \gamma_{ISB}$, the reduction in the density of active electrons can be important already at small excitation levels $\Delta E/E_{21}^o\ll 1$. This suggests that in quasi-CW illumination regimes the nonlinear shift of the resonances receives a dominant contribution from incoherent saturation effects. If one is interested in incoherent nonlinear processes such as bleaching and/or a frequency shift of the ISB transition, the long relaxation time $\gamma_d$ is a beneficial feature to reduce the required incident power~\cite{colombelli2020absorption}.

In spite of the presence of incoherent effects, the coherent nonlinearities that underlie  wave-mixing effects remain however active and display an interestingly different scaling with $\sigma_{el}$: given the form of the last, nonlinear term of \eqref{eq:shift_omega}, the excitation level $\Delta E/E_{21}^o$ needed to obtain a given value of the (coherent) nonlinear shift decreases as $\sigma_{el}^{-1}$ for growing electron density $\sigma_{el}$ for fixed QW geometry. Via \eqref{eq:el_act}, this implies that the incoherent saturation effect can be reduced by increasing the electron density $\sigma_{el}$. Furthermore, under the experimentally reasonable assumption that $\gamma_d$ does not change much when moving from the Mid-IR towards the THz range, this same equation suggests that the relative effect of the incoherent nonlinearities is reduced  for longer wavelengths. 

%Based on the experimental observation that $\gamma_d$ is approximately constant across the different wavelength ranges while $\gamma_X$ scales with the resonance frequency, one might anticipate that the relative importance of the incoherent nonlinearity decreases upon moving from Mid-IR towards the THz range.
%In spite of these incoherent effects, coherent nonlinearities are still active and can be used for exciting applications but one needs to pay attention to the additional complexities of the dynamics and to the interpretation of the observations. 
%On one hand, the actual strength of the coherent nonlinearities can not be straightforwardly extracted from the frequency shift of the resonances under quasi-CW excitation as measured, e.g., in pump-and-probe experiments with a quasi-CW pump~\cite{Hawecker}. The observed frequency shift gets in fact contributions from both coherent and incoherent nonlinearities and does not allow to separate the two contributions.

Finally, it is important to note that all these arguments hold a continuous-wave illumination of the sample by, e.g., a quantum cascade laser source~\cite{Hawecker}. A promising alternative to further suppress the incoherent effects is to use a pulsed excitation. In this regime, the experiment can be carried out on a fast enough time scale that the interesting coherent nonlinear dynamics occurs before a sizable amount of dark excitations is generated~\cite{Knorr}.

\section{Optical nonlinearities in microcavities}
\label{sec:microcav}

One of the most promising configurations to exploit the optical nonlinearities discussed in the previous Section to observe useful optical processes is to embed the QWs within high-$Q$ microcavity devices so to enter the so-called strong light-matter coupling regime. In this regime, the  elementary  excitation  modes  have  the  mixed light-matter character of polaritons, which allows for an efficient coupling of the electronic degrees of freedom to the optical fields and therefore enhances the effect of the nonlinearities.
Building a complete theory of the nonlinear dynamics of such microcavity devices is a task that goes far beyond this work for which preliminary steps have been reported in Ref.~\onlinecite{nespolo2019generalized,colombelli2020absorption,Knorr}. As such, the goal of this Section is to obtain a quantitative estimate of the actual strength of the ISB nonlinearities in a configuration that is most promising for applications.

As a specific benchmark quantity, we will consider the light intensity value that is needed to have a frequency shift of the polariton mode comparable to the linewidth. This is the typical condition under which important nonlinear effects such as optical bistability~\cite{carusotto_quantum_2013} or optical parametric oscillation start occurring~\cite{nespolo2019generalized}. The discussion that follows will mostly concentrate on the latter effect, which is a promising strategy to achieve lasing and Bose-Einstein condensation effects in novel wavelength regions~\cite{Manceau_PRX15,bloch2021spontaneous}. Since the amplitude of the parametric coupling between the pump and the signal/idler modes is quantitatively related to the frequency shift, it is natural to characterize the parametric oscillation threshold in terms of the ratio between the frequency shift and the decay rate. 

Even though our theory is fully general and can be applied to generic devices, for the sake of concreteness we keep in mind the specific example of double-metal microcavities, a most promising work-horse for studies of ISB polaritons in both the THz and mid-infrared frequency ranges~\cite{manceau-critical}. 
In these devices, 
the QWs are sandwiched between two metallic layers. The back metallic layer is left unpatterned and acts as a perfectly reflecting plane, while the front one is periodically patterned to allow optical access from the far field. Owing to this {\em single-sided} geometry, all spectroscopic information can be obtained from reflectivity measurements since there is no transmitted beam. Furthermore, the efficiency of the nonlinear process can be optimized by independently tailoring the different decay channels~\cite{manceau-critical,Todorov-THz-enhanc} 
so to reach the so-called {\em critical coupling} regime with external radiation, where radiative and non-radiative losses are equal and on resonance all incident light is funneled into the cavity~\cite{zanotto2014perfect}.

Under this condition, the energy density stored in the cavity is simply related to the incident power $P_{inc}$ by $\varepsilon_{st}\,\gamma_{\rm pol}=P_{inc}$ where $\gamma_{\rm pol}$ is the polariton decay rate.
In the strong-coupling regime, the energy of a polariton mode is shared by its light and matter component in proportion to the Hopfield coefficients, so the excitation density in each well and per electron is given by
\begin{equation}
\sigma_{el} \Delta E= \frac{|u_X|^2\,\varepsilon_{st}}{N_w} = \frac{P_{inc}\, |u_{X}|^2}{ N_w \gamma_{\rm pol} }
\end{equation}
where $N_w$ is the number of QWs coupled to the cavity mode and $|u_X|^2$ is Hopfield coefficient quantifying the matter component of the polariton.

\subsection{Nonlinearity from the depolarization shift}

Deep in the strong coupling, the frequency shift of the polariton mode is $|u_X|^2$ times the one of the matter excitation~\cite{carusotto_quantum_2013}. This result can be combined with the analytical formula \eqref{eq:shift_omega} for the ISB frequency shift to obtain an explicit expression for the power-dependent frequency-shift of the polariton,
\begin{equation}
    \Delta (\hbar\omega_{\rm res})= -a_1 \frac{e^2 L_{QW}}{\epsilon} \frac{|u_X|^4}{N_w \gamma_{\rm pol} E_{21}^o}\,P_{inc}\,.
\end{equation}
From this formula, assuming for simplicity $|u_X|^2=1/2$ and a typical number $N_w=10$ of wells, we can estimate that an intensity around
$1\,\textrm{MW}/\textrm{cm}^2$ is required to obtain a red-shift of the polariton mode comparable to the polariton linewidth $\hbar \gamma_{\rm pol}=5$~meV. Quite interestingly, note that this formula does not involve the electron density. This is of course valid as long as one remains in the strong coupling regime~\cite{colombelli2020absorption}. In contrast, as we have pointed out in Sec.\ref{sec:incoh}, the incoherent saturation effect  at a given value of the nonlinear frequency shift is smaller for a large electron density.

Based on the scaling laws discussed above, 
and plugging in the typical experimental observation that the quantity $\hbar\gamma_{\rm pol}/E_{21}^o$ (that is, the $Q$ factor) is typically constant across the different frequency windows, one obtains that that the required power to achieve a red-shift comparable to $\hbar \gamma_{\rm pol}$ scales as $\alpha^{7/2}$ and thus quickly decreases as one moves to longer wavelengths. 
As a concrete example, reducing the resonance frequency by a factor $10$ from $110$~meV to $11$~meV (corresponding to $\sim 2.7$~THz) reduces the $P_{inc}$ by a remarkable factor $\sim 3000$ towards the few $100\,\textrm{W}/\textrm{cm}^2$ range. Further reductions could come from a reduction of the number $N_w$ of wells (keeping a fixed overlap factor), an improvement of the cavity $Q$-factor, or a clever design of the cavity so to spatially concentrate the light intensity in subwavelength volumes~\cite{Paulillo_2014,Todorov_2015}.

\subsection{Nonlinearity from the saturation of polariton splitting} 
When dealing with microcavity configurations, it is important to remind that an additional frequency shift of the polariton modes arises from the nonlinear saturation of the dipole moment which induces a corresponding reduction of the polariton  Rabi splitting~\cite{carusotto_quantum_2013}. 

Within our theory, this effect is captured by the nonlinear dependence of the dipole moment in \eqref{eq:z0}, which gives a corresponding variation of the oscillator strength, $f\simeq f_0 \, (1-b_1\,\Delta E/E_{21}^o)$. In terms of the polariton splitting $\Omega_R$, this results into
\begin{equation}
    \Omega_R =\Omega_R^o +\Delta \Omega_R \simeq \Omega_R^o \, \left(1 - b_1 \frac{\Delta E}{E_{21}^o} \right)\,.
\end{equation}
where the linear-regime Rabi frequency is given by
\begin{equation}
    \Omega_R^o=\left(\frac{2\pi e^2 N_w \sigma_{el}\eta}{\epsilon \, m^* L_{cav}}\right)^{1/2}\,.
\end{equation}
Here, $\eta$ is an adimensional parameter of geometric origin, typically of order one, while $L_{cav}$ is the  thickness of the cavity.

It is interesting to quantitatively compare the magnitude of the nonlinear shift due to the ISB frequency shift \eqref{eq:shift_omega} to the one coming from this reduction of $\Omega_R$. To this purpose, we can consider the ratio
\begin{equation}
    \frac{\Delta \Omega_R}{\Delta(\hbar \omega_{res})}=\frac{b_1}{a_1} \frac{\Omega_R}{E_{21}^o}\,\frac{\epsilon E_{21}^o}{e^2 \sigma_{el} L_{QW}}\,.
    \label{eq:nonl_ratio}
\end{equation}
Plugging in the specific parameters for the Mid-IR QW considered above with an electron density $\sigma_{el}=3\cdot 10^{12}\,\textrm{cm}^{-2}$, a QW density $N_w/L_{cav}=0.02\,\textrm{nm}^{-1}$ and $\eta=1/2$, one obtains $\Omega_R^o \simeq 16\,\textrm{meV}$. Inserting this value into \eqref{eq:nonl_ratio}, one finds that the ratio of the two nonlinearities is in the order of unity.
Interestingly, the last factor on the RHS of \eqref{eq:nonl_ratio} is constant under our usual scaling while, for a given overlap factor $N_w/L_{cav}$, the ratio $\Omega_R^o/E_{21}^o$ displays a slow variation as $\alpha^{-1/4}$. As a result, one can not expect major changes in this ratio when moving from Mid-IR towards the THz range. This confirms that our arguments on the scaling of the required incident power with operation wavelength remain valid when we include this saturation nonlinearity.

Note that a very different behaviour is expected for the two polariton branches. For the upper polariton, both mechanisms give rise to a red-shift and cooperate to reinforce the nonlinear effect. For the lower polariton, instead, they push in opposite directions and, depending on the actual value of the Hopfield coefficients, they may cancel out, suppressing the final value of the effective nonlinearity. These arguments suggest that in the ISB case the upper polariton branch is more favourable for nonlinear optics experiments. This conclusion~\footnote{Note that a positive interaction coefficient between ISB polaritons $g>0$ was assumed by two of us in Ref.~\onlinecite{nespolo2019generalized}. This erroneous choice was motivated by the analogy with the exciton-polariton case and, {\em mutatis mutandis} does not bring major prejudice to the overall conclusions of that work.}
is to be contrasted with the exciton-polariton case where the frequency shift of the exciton under the effect of the repulsive binary interactions is in the blue direction, making the lower polariton a more favourable choice for nonlinear optics experiments~\cite{carusotto_quantum_2013}.

\section{Effective quantum Hamiltonian}
\label{sec:quantum}

Even though the theory presented in this work is based on the excitation of the electronic system by classical light, our results are a good starting point to attempt a phenomenological quantum theory of optical nonlinearities of electrons in QWs. Such a development is of utmost importance if one is to extend quantum optics concepts, tools and applications originally developed for visible or near-IR light~\cite{walls2007quantum} to devices operating in longer wavelength ranges. Building a complete theory of quantum nonlinearities is a task that goes way beyond this work, so we will restrict here to some semi-quantitative reasonings that offer an intuitive feeling of the strength of the effect.

Indicating with $\hat{\Psi}_{\rm X}(\mathbf{r})$ the (approximately bosonic) field operator describing the bright ISB excitation mode of the electrons in the QW~\cite{Bastard2005,todorov_intersubband_2012},% {\bf add some reference on bright/dark states}, 
we can write a model Hamiltonian in the form:
\begin{multline}
 \mathcal{H}=\hbar\omega_{\rm lin}\,\int\! d^2\mathbf{r}\, \hat{\Psi}^\dagger_{\rm X}(\mathbf{r})\,\hat{\Psi}_{\rm X}(\mathbf{r}) + \\ + \frac{\hbar \omega_{\rm nl}}{2}\,\,\int\! d^2\mathbf{r}\, \hat{\Psi}^\dagger_{\rm X}(\mathbf{r})\,\hat{\Psi}^\dagger_{\rm X}(\mathbf{r})\,\hat{\Psi}_{\rm X}(\mathbf{r})\,\hat{\Psi}_{\rm X}(\mathbf{r}) + \\ - \left. e \,d_0\int\!d^2\mathbf{r}\,\mathcal{E}(\mathbf{r},t)\hat{\Psi}_{\rm X}^\dagger(\mathbf{r})\left(1- \frac{\hat{\Psi}^\dagger_{\rm X}(\mathbf{r})\,\hat{\Psi}_{\rm X}(\mathbf{r})}{n_{\rm sat}}\right) \right.+ \\ \left. - \textrm{h.c.}\right.
 \label{eq:Hquantum}
\end{multline}
where %{\bf faire le lien avec le $g$ dans GPE de JN}
$\hbar\omega_{\rm lin}=E_{21}^o+a_0 \, e^2 L_{QW}/\epsilon$ is the linear oscillation frequency and the binary interaction energy $\hbar\omega_{\rm nl}=-
a_1\, e^2 L_{QW}/\epsilon$
accounts for the red-shift of the resonance. Note that this effective interaction term has an opposite sign compared to the exciton-polariton case~\cite{carusotto_quantum_2013}. 

The term describing the coupling to the applied electric field $\mathcal{E}(\mathbf{r},t)$ has the physical meaning of an effective transition dipole. At linear regime its value is
\begin{equation}
    d_0=b_0 L_{QW} \,\left[1-s_1\frac{e^2 \sigma_{el} L_{QW}}{\epsilon E_{21}^o}\right]\,\sqrt{\sigma_{el}},
\end{equation}
while at higher densities displays a saturation behaviour of saturation density $n_{sat}=\sigma_{el}/b_1$. As usual, the operator
$\hat{\Psi}^\dagger_{\rm X}(\mathbf{r})\,\hat{\Psi}_{\rm X}(\mathbf{r})$ indicates the in-plane density of ISB excitation quanta. In the language of Ref.~\onlinecite{Todorov:PRX14}, our quantum Hamiltonian \eqref{eq:Hquantum} refers to the ``bosonic'' regime of a relatively large number of electrons and relatively weak excitation.

As a simplest example of application of this model Hamiltonian, it is interesting to estimate the strength of the single-excitation nonlinearity in a subwavelength resonator of lateral area $S_{\rm cav}$ where the electromagnetic field is confined in all three-dimensions~\cite{Paulillo_2016,keller2017few,Jeannin2019ultrastrong,malerba2016}. %{\bf C2N: add more ref's on different cavities if needed}. 
As a figure of merit, we will consider the frequency shift $\Delta_1$ of the polariton resonance when a single quantum of excitation is injected into the device. When $\Delta_1$ exceeds the linewidth $\gamma_{\rm pol}$, the presence of a single quantum of excitation is able to push the oscillation frequency away from resonance with the incident light and, in this way, prevent the injection of a second quantum of energy into the device. This phenomenon goes under the name of photon/polariton blockade and is experimentally visible as strong non-classical features in the transmitted and reflected light such as antibunching~\cite{imamoglu_blockade,carusotto_quantum_2013,verger2006polariton}.

In order to estimate $\Delta_1$, we first note that the energy of a single quantum of excitation will distribute among the $N_{\rm el}=\sigma_{el}S_{\rm cav}$ electrons present in the resonator, giving an excitation density $\Delta E/E_{21}^o = 1/N_{\rm el}$. As a result,
\begin{equation}
    \frac{\Delta_1}{\gamma_{\rm pol}} \simeq \frac{a_1 e^2 L_{QW} \sigma_{el}}{\epsilon\,E_{21}^o} \, \frac{E_{21}^o}{\hbar \gamma_{\rm pol}} \frac{1}{N_{\rm el}}\,.
    \label{eq:patch}
\end{equation}
For the Mid-IR configuration considered in this work, the first fraction on the RHS is of order $0.4$, so blockade $\Delta_1/\gamma_{\rm pol} \geq 1$ requires the number of electrons $N_{\rm el}$ to be a sizable factor below the $Q$ factor of the cavity, indicated here by the $E_{21}^o/(\hbar \gamma_{\rm pol})$ factor. Assuming that electrons are uniformly distributed in the cavity area with a given two-dimensional density, this imposes an upper bound on the cavity area.

As a quantitative benchmark, for the electron density value $\sigma_{el}=3\cdot 10^{12}\,\textrm{cm}^{-2}$ used so far, each electron effectively occupies a region of $\sim 30\,\textrm{nm}^2$. Using sub-wavelength nano-antennas and comparable electron densities, it was possible to achieve a lateral confinement of the field strong enough to observe strong light-matter coupling in the mid-infrared with a few $10^3$ electrons confined in an area of a characteristic linear size of $100\,$nm~\cite{malerba2016}. 

Based on our scaling arguments for the different coefficients in \eqref{eq:patch}, it is immediate to see that the criterion based on the $Q$ factor and the number $N_{\rm el}$ of electrons directly extends to longer wavelength regimes and leads to comparable if not more promising predictions for THz radiation. The larger dipole moment of the transition reduces in fact the required electronic density for strong coupling and thus weakens the constraint on the maximum physical size of the patch cavity to observe blockade. As a result, ultra-strong light-matter coupling has been observed using a cyclotron transition of less than $100$ electrons coupled to a sub-THz nanogap hybrid LC microcavities~\cite{keller2017few} and using a $3$~THz ISB transition of a few $1000$ electrons coupled to a LC resonator~\cite{Jeannin2019ultrastrong, goulain_thz_2023}. 

%{\bf citer papier Tavis-Cummings de Yanko~\cite{Todorov:PRX14}}
Comparing these values with the $Q$ factors in the $20$ range that are presently available and considering the perspectives of further improvement sketched in the original works, these results are extremely promising in view of reaching polariton blockade in the Mid-IR and THz domains in the next future. Given the present state of technology, it is likely that a main experimental hurdle along this path will consist of the development of efficient single photon detectors to measure quantum correlations for such long-wavelength radiation.

\section{Conclusions and perspectives}
\label{sec:conclu}
To summarize, in this work we have developed a general theory of the coherent optical nonlinearities associated to intersubband transitions in semiconductor quantum wells including the quantum mechanical motion of electrons and their Coulomb interactions. As most relevant observable quantities, simple expressions for the intensity-dependence of the oscillation frequency and the dipole moment of the intersubband transition are derived. Interesting scaling laws in the operation wavelength are highlighted and crucial differences from competing processes such as incoherent saturation effects are pointed out. The consequences of these optical nonlinearities on intersubband polaritons in microcavity geometries are investigated and quantitative estimates across different ranges of wavelengths from the Mid-IR to the THz are put forward. These predictions appear promising in view of the observation of novel phenomena such as parametric gain and Bose-Einstein condensation of intersubband polaritons. Finally, as a more speculative direction, we have explored the potential of intersubband polaritons as a platform for exploting blockade effects to generate antibunched light in longer wavelength ranges where quantum optics is still much less developed. 

The theoretical framework discussed in this work will be of great use in future work to design structures with more complex potentials, so to maximize the strength of the nonlinear response for different processes such as intensity-dependent frequency shifts, harmonic generation, and coherent wave-mixing processes. On the longer run, our results will be a useful starting point to build a fully quantum optical theory to guide experiments aiming at extending, e.g., photon blockade phenomena and single-photon emission to novel wavelength regimes. 

%\acknowledgments
\begin{acknowledgements}
We acknowledge financial support from the European Union FET-Open grant MIR-BOSE (737017). The Trento team acknowledges support from the Provincia Autonoma di Trento. Continuous stimulating discussions with J. Hawecker, S. Dhillon, M. Knorr, C. Lange, R. Huber, A. Tredicucci, G. Scalari, J. Faist are warmly acknowledged.
\end{acknowledgements}

\appendix

\section{Imaginary-time evolution
\label{app:ImT}}
In this Appendix, we briefly review the main principles underlying the imaginary-time evolution method used to find the wavefunction $\psi_g(z)$ and the energy $E_g$ of the lowest-energy eigenstate of the time-independent Schr\"odinger-Poisson equation \eqref{eq:NLNLSE_stat}.
%  \begin{multline}
% E_g\,\psi_g(z)=
% -\frac{\hbar^2}{2m^*}\frac{\partial^2 \psi_g}{\partial z^2} +
% V_{QW}(z)\psi_1(z) + \\
% +\frac{2\pi\, e^2}{\epsilon} \int\!dz'\,|z-z'|\,\left[\sigma_{imp}(z')-\sigma_{el}|\psi_1(z')|^2\right]\,\psi_1(z)\,.
% \label{eq:SP0}
% \end{multline}

The partial differential equation encoding the imaginary-time evolution is obtained from the real-time evolution equation \eqref{eq:NLNLSE} by rotating the time variable $t\rightarrow -i\beta$ in the complex plane, 
 \begin{multline}
\frac{\partial\psi(z,\beta)}{\partial \beta}=-\frac{1}{\hbar}\left\{
-\frac{\hbar^2}{2m^*}\frac{\partial^2 \psi}{\partial z^2} +
V_{QW}(z)\psi(z) +\right. \\
\left.+\frac{2\pi\, e^2}{\epsilon} \int\!dz'\,|z-z'|\,\left[\sigma_{imp}(z')-\sigma_{el}\frac{|\psi(z')|^2}{|| \psi ||^2}\right]\,\psi(z)\right\}.
\label{eq:NLNLSE_ImT}
\end{multline}
The factor involving the norm of the wavefunction 
\begin{equation}
    || \psi ||^2 = \int \!dz\, |\psi(z)|^2
\end{equation}
is required at the denominator of the interaction term since the imaginary-time evolution (in contrast to the real-time one) does not conserve the norm.

At long times $\beta\to\infty$, the imaginary-time evolution typically converges to an exponentially decreasing wavefunction \begin{equation}
    \psi(z,\beta)\simeq \psi_\infty(z)\,e^{-\beta E_\infty}
    \label{eq:limit_beta}
\end{equation} 
from which one extracts the ground state wavefunction $\psi_g(z)=\psi_\infty(z)/||\psi_\infty||$ and the ground state energy $E_g=E_\infty$. Inserting the ansatz \eqref{eq:limit_beta} into \eqref{eq:NLNLSE}, one indeed recovers the time-independent SP equation of the form \eqref{eq:NLNLSE_stat}.

In a practical calculation, we can choose a generic wavefunction as the initial state $\psi(z,\beta=0)$. We then have to numerically evolve $\psi(z,\beta)$ in $\beta$ according to \eqref{eq:NLNLSE_ImT} until we reach convergence. This is determined by looking at the convergence of the SP energy of $\psi(z,\beta)$ to a constant value. The imaginary-time evolution (as well as the following real-time one) is carried out using a split-step method: the evolution at each time-step is Trotter-split into the non-commuting kinetic and potential energy parts and each of them is sequentially implemented in the space in which it is diagonal, namely $k$-space for the kinetic energy and real-space for the potential and interaction energy terms. At each time-step, inter-conversion between the $k$- and the real-space and back is performed by Fast Fourier Transform. 

The imaginary-time method is  most transparent in the non-interacting limit where the nonlinear term in the evolution equation is negligible. In this case, we can decompose the initial wavefunction 
\begin{equation}
    \psi(z,\beta=0)=\sum_j a_j(\beta=0)\,\psi_j(z)
\end{equation}
on the orthonormal basis of eigenfunctions $\psi^o_{j=1,2,\ldots}(z)$ of the Schr\"odinger problem in the bare quantum well potential, with energy $E^o_{j=1,2,\ldots}$. The random initial condition reflects into a random choice of the initial value $a_j(\beta=0)$ of the expansion coefficients.
By linearity, the imaginary-time evolution acts independently on each of them, $a_j(\beta)=a_j(0)\,e^{-\beta E^o_j }$. At late times $\beta\to \infty$, only the lowest-energy eigenvector survives (the higher ones $j\geq 1$ are exponentially suppressed at least as $e^{-\beta(E^o_2-E^o_1)}$) and $\psi(z,\beta)$ converges to the lowest-energy eigenvector,
\begin{equation}
 \psi(z,\beta)\simeq a_1 e^{-\beta E^o_1 }\,\psi^o_1(z)   
\end{equation}
recovering the limiting form \eqref{eq:limit_beta}. 

From this discussion, it is immediate to see that the correct ground state is found independently on the choice of the initial wavefunction provided $a_1(\beta=0)\neq 0$, a condition which is satisfied by any randomly-chosen initial wavefunction. This independence from the initial condition, mathematically proven in the non-interacting case, has been numerically verified to also hold in the interacting case by repeating the calculation for different choices of the initial wavefunction.

\section{Analytical study of the oscillation frequency in the limit of low electron density}
\label{app:Bogo}

As a further verification of the numerical calculations and a quick guiding tool for the design of new structures, it is interesting to look at the linearized equation \eqref{eq:Bogo_t} in the small electron density limit where an analytical treatment is possible. The discussion in this Appendix is inspired from the Bogoliubov theory of the weakly interacting Bose gas~\cite{pitaevskii2016bose}. 

We indicate with $\psi^o_{1,2}(z)$ the two lowest states of the quantum well for negligible electron density.
In the limit of a weak excitation, we can expand the linearized dynamics in the basis of these two states only, 
\begin{multline}
\psi(z,t)=e^{-iE_1^o t/\hbar}  \times \\ \times \left[\psi^o_1(z) + \alpha(t)\,\psi^o_2(z)\, u_2 + \alpha^*(t)\,\psi^o_2(z)\,v_2  \right]
\label{eq:Bogo_ansatz}
\end{multline}
where $\alpha(t)$ is the excitation amplitude and $[u_2,v_2]^T$ is the projection of the linearized eigenmode $[u(z),v(z)]^T$ on the excited $\psi_2$ state to which we are restricting our attention, with the usual normalization $|u_2|^2-|v_2|^2=1$.

Plugging the ansatz \eqref{eq:Bogo_ansatz} into the linearized SP dynamics \eqref{eq:Bogo_t} and imposing a harmonic evolution of the excitation amplitude, $\alpha(t)=\bar{\alpha}\,e^{-i\omega_{\rm res} t}$, we get to the eigenvalue equation,
\begin{equation}
\mathcal{L} \left(\begin{array}{c} u_2 \\ v_2 \end{array} \right)= \hbar \omega_{\rm res} \left(\begin{array}{c} u_2 \\ v_2 \end{array} \right)
\end{equation}
with
\begin{equation}
\mathcal{L} = \left(\begin{array}{cc} E^o_{21} + \Delta_H + \Delta_x %(V_{eg} -V_{gg}+V_x) 
& \Delta_x \\ -\Delta_x & -E^o_{21} - \Delta_H - \Delta_x
%-\hbar\omega_eg - \eta (V_{eg} -V_{gg}+ V_x) 
\end{array} \right)\,.
\label{eq:Bogo_matrix}
\end{equation}
Here, $E^o_{21}=E^o_2-E_1^o$ is the energy difference between bare electronic levels. 
The terms accounting for the Coulomb interactions are proportional to the dimensional ${\bar \eta}=2\pi e^2 \sigma_{el}/\epsilon$ coefficient quantifying the effective strength of Coulomb interactions. In detail, 
\begin{equation}
\Delta_H=V_{21}-V_{11}
\label{eq:Hartree}
\end{equation}
is the static Hartree shift of the transition under the effect of the Coulomb interactions, expressed in terms of the static Coulomb shift of the $\psi_{1,2}$ states in the charge distribution determined by the ground state electrons and the impurities,
\begin{eqnarray}
V_{11}&=&-{\bar \eta} \, \int\! dz \! \int\! dz'\, |z-z'|\left[ |\psi^o_1(z)|^2 - \sigma_{imp}(z)\right] \, |\psi^o_1(z')|^2  \nonumber \\ \, \label{eq:Bogo1}\\
V_{21}&=&-{\bar \eta} \, \int \!dz\!\int \!dz'\, |z-z'|\left[ |\psi^o_1(z)|^2 - \sigma_{imp}(z)\right] \, |\psi^o_2(z')|^2 \nonumber \\ \, \label{eq:Bogo2}
\end{eqnarray}
and $\Delta_x$ accounts for the dynamical distortion of the electronic distribution induced by the excitation,
\begin{equation}
\Delta_x= - {\bar \eta} \, \int\! dz\!\int \!dz'\, |z-z'|\, \psi^o_1(z)\,\psi^o_2(z)\,\psi^o_1(z')\,\psi^o_2(z')\,. \label{eq:Bogo3}
\end{equation}
For low electron densities, all these quantities are small compared to the bare transition energy $E_{21}^o$, so the resonance energy $\hbar E_{\rm res}$ can be analytically calculated within perturbation theory as 
\begin{equation}
\hbar \omega_{\rm res} \simeq E^o_{21} + \Delta_H + \Delta_x - \frac{\Delta_x^2}{2E^o_{21}}\,.
\label{eq:omegares}
\end{equation}

\begin{table}
\begin{tabular}{|c|c|c|c|}
\hline
 & \parbox[c]{1.75cm}{outside \\ $\pm 10$\,nm} & \parbox[c]{1.75cm}{inside \\ $\pm 2\,$nm} & \parbox[c]{1.75cm}{center } \\ [1ex] \hline
 $V_{11}/{\bar \eta}$ &  0.79 & -0.011 & -0.06 \\
 $V_{21}/{\bar \eta}$ &  0.73 & -0.013 & -0.02 \\
 $\Delta_{H}/{\bar \eta}$ &  -0.06 & -0.002 & 0.04 \\
 $\Delta_{x}/{\bar \eta}$ & 0.11 & 0.11 & 0.11 \\ \hline
$\hbar \omega_{\rm res}-E_{21}$  & 9.5\,meV & 26.6\,meV & 37.7\,meV \\
$\Delta_H$  & -17.2\,meV & -0.55\,meV & 10.3\,meV  \\
\hline
\end{tabular}
\caption{\label{tab:integrals} Table of the Coulomb interaction integrals (\ref{eq:Bogo1}-\ref{eq:Bogo3}) and of the energy shifts \eq{eq:Hartree} and \eq{eq:omegares} for an infinite well of thickness $L_w=12.7$\,nm for which the transition energy $\Delta_{eg}$=109.6\,meV is comparable to the case considered in the main text. The different columns refer to different locations of the impurities, namely outside the well [as in the left panel of Fig.\ref{fig:fig_algo}(b)], inside the well [as in the right panel of Fig.\ref{fig:fig_algo}(b)], at the center of the well. The energy shifts on the last two rows are evaluated for an electronic density  $\sigma=3\cdot 10^{12}\,\textrm{cm}^{-2}$ corresponding to the rightmost points in Fig.\ref{fig:fig_linear}(b).}
%\multicolumn{4}{|c|}{\small{(for $\sigma=3\cdot 10^{12}\,\textrm{cm}^{-2}$)}} \\ \hline}
\end{table}

These quantities can be easily computed in the idealized case of an infinite well of thickness $L_w$ for which the wavefunctions and the energies of the two lowest states and their energies have the following analytical forms,
\begin{eqnarray}
\psi^o_1(z)&=& \sqrt{\frac{2}{L_w}}\,\cos\left(\frac{\pi z}{L_w}\right), \hspace{0.105cm} \hbar E^o_1=\frac{\hbar^2}{2m^*}\left(\frac{\pi}{L}\right)^2  \\
\psi^o_2(z)&=&\sqrt{\frac{2}{L_w}}\,\sin\left(\frac{2 \pi z}{L_w}\right), \hspace{0.105cm} E^o_2=\frac{\hbar^2}{2m^*}\left(\frac{2\pi}{L}\right)^2\,.
\end{eqnarray}
In particular, the value of the integrals in (\ref{eq:Bogo1}-\ref{eq:Bogo3}) can be straightforwardly estimated for different locations of the impurities as summarized in the upper part of Table \ref{tab:integrals}.

Inserting these values in the matrix \eqref{eq:Bogo_matrix}, one obtains the prediction for the shift of the ISB transition listed in the bottom part of Table \ref{tab:integrals}. In spite of the severe approximations done, this prediction is in quite good quantitative agreement with the full numerical calculation shown in Fig.\ref{fig:energyres}(b). Note how the resonance shift strongly departs from the static Hartree prediction $\Delta_H$ based on the static Coulomb energy shift of single-particle orbitals.

It is interesting to remind that this crucial difference between the single-particle energies and the collective excitation frequencies also occurs in dilute Bose gas with local interactions~\cite{pitaevskii2016bose,castin2001bose}. For instance, in a spatially homogeneous geometry, the single-particle states have the quadratic dispersion of the kinetic energy, while the Bogoliubov dispersion of the collective excitation modes obtained from the linearization of the GPE starts at low wavevector with a sonic dispersion dominated by the interaction effects and only later recovers the quadratic form.

\bibliography{nonlinresponse,mybib}

\end{document}